\documentclass[a4paper,11pt]{article}
\usepackage{pos}
\usepackage{slashed}

\title{
Nucleon mass in covariant baryon chiral perturbation theory at leading two-loop order
}
\author*[a]{Ze-Rui Liang}
\author[b,c]{Han-Xue~Chen}
\author[d,e]{Feng-Kun~Guo}
\author[a]{Zhi-Hui~Guo}
\author[b,c]{De-Liang~Yao}

\affiliation[a]{
College of Physics and Hebei Key Laboratory of Photophysics Research and Application, Hebei Normal University, \\ 
Shijiazhuang, Hebei 050024, China
}

\affiliation[b]{
School of Physics and Electronics, Hunan University, \\
Changsha 410082, China
}

\affiliation[c]{
Hunan Provincial Key Laboratory of High-Energy Scale Physics and Applications, Hunan University, \\
Changsha 410082, China
}

\affiliation[d]{
Institute of Theoretical Physics, Chinese Academy of Sciences,  \\
Beijing 100190, China
}

\affiliation[e]{
Peng Huanwu Collaborative Center for Research and Education, Beihang University, \\
Beijing 100191, China
}

\emailAdd{liangzr@hebtu.edu.cn}
\emailAdd{chenhanxue@hnu.edu.cn}
\emailAdd{fkguo@itp.ac.cn}
\emailAdd{zhguo@mail.hebtu.edu.cn}
\emailAdd{yaodeliang@hnu.edu.cn}

\abstract{

We calculate the nucleon mass within a manifestly relativistic formulation of baryon chiral perturbation theory (BChPT), extending the framework to leading two-loop order ($\mathcal{O}(p^5)$). By employing dimensional counting analysis and rigorously verifying the extended on-mass-shell scheme at this order, we obtain a complete chiral representation of the nucleon mass that preserves analyticity, respects proper power counting, and maintains renormalization-scale independence. The resulting expression exhibits excellent convergence, with $\mathcal{O}(p^5)$ contributions remaining small ($\sim 8~\rm{MeV}$). This formulation provides a robust foundation for chiral extrapolation, demonstrating remarkable agreement with lattice QCD data across a wide range of pion masses ($M_\pi \lesssim 300~\rm{MeV}$). The success of this calculation establishes two-loop relativistic BChPT as a precision tool for studying nucleon structure and related properties.

}

\FullConference{The 21st International Conference on Hadron Spectroscopy and Structure (HADRON2025)\\
 27 - 31 March, 2025\\
Osaka University, Japan\\}

\begin{document}
\maketitle

\section{Introduction}

The origin of nucleon mass is particularly important for our understanding of the low-energy QCD dynamics.   
The decomposition of nucleon mass can be achieved via the matrix element of the trace of the energy-momentum tensor $T^\mu_\mu$~\cite{Ji:1994av,Lorce:2017xzd},
\begin{align}
m_N=\langle N(p)|T^\mu_\mu |N(p) \rangle \sim\langle N(p) | \underbrace{\frac{\beta}{2g} G_{\mu\nu}^aG_a^{\mu\nu}}_\text{field energy}+\underbrace{m_u\bar{u}u+m_d\bar{d}d+m_s\bar{s}{s}}_\text{Higgs}|N(p)\rangle \ .
\end{align}
The first term is the trace anomaly responsible for the gluonic contribution to the nucleon mass, and the second term is the quark mass term. The quark mass term is also called the nucleon sigma term, whose value quantifies the amount of the quark mass contribution to the nucleon mass. The sigma term characterizes the scalar coupling of the nucleon and is therefore relevant to the dark matter detections~\cite{Bottino:1999ei}. The nucleon sigma term can be derived from the nucleon mass by virtue of the Feynman-Hellman theorem~\cite{Hellmann:1937book,Feynman:1939zza}. A precise determination of the pion-nucleon sigma term requires the nucleon mass with high-order chiral corrections beyond one loop.

In BChPT, there have been many works devoted to high-order calculations of the nucleon mass. In 1999, the nucleon mass was first calculated in the heavy baryon formalism up to $\mathcal{O}(p^5)$, corresponding to the leading two-loop order~\cite{McGovern:1998tm}. Later, the chiral correction to the nucleon mass was computed  up to $\mathcal{O}(p^6)$, namely, the complete two-loop order~\cite{Schindler:2006ha,Schindler:2007dr}, by using a relativistic version of BChPT with infrared regularization (IR) prescription~\cite{Becher:1999he}. It is only until 2024 that the extended on-mass-shell (EOMS) scheme was applied to study the nucleon mass at two-loop order~\cite{Conrad:2024phd,Conrad:2024sla,Chen:2024twu}. Nevertheless, in Refs.~\cite{Conrad:2024phd,Conrad:2024sla,Chen:2024twu}, the result of the nucleon mass is expanded in powers of the pion mass, and the full EOMS result is absent. Our work focuses on deriving a full chiral expression for the nucleon mass at leading two-loop order~\cite{Liang:2025cjd}, which respects the correct power counting (PC) and preserves the original analyticity. Note that the EOMS scheme has been extensively used to investigate various quantities at one-loop level, see e.g. Refs.~\cite{Ren:2016aeo, Yao:2016vbz,Liang:2023scp}. Nowadays, its extension to two-loop calculations is required to describe modern data from lattice QCD and experiments with high precision. 

\section{Leading two-loop nucleon mass in BChPT\label{sec2}}
The definition of the nucleon mass $m_N$ is given by the pole of the full nucleon propagator, i.e.,
\begin{align}
iS_N(\slashed{p})=\frac{i}{\slashed{p}-m-\Sigma_N(\slashed{p})}=\frac{i Z_N}{\slashed{p}-m_N}+\text{non-pole~piece} \ ,
\end{align}
where the residue $Z_N$ is the wave renormalization constant and $\Sigma_N(\slashed{p})$ is the self-energy of the nucleon. Specifically, the nucleon mass is obtained by solving the following equation, 
\begin{align}
m_N-m-\Sigma(m_N,m)=0\ .
\end{align}
For the sake of easy explanation, we can express the nucleon mass in the form as
\begin{align}
m_N=m_4+\hbar \Delta m_N^{(1)} + \hbar^2\Delta m_N^{(2)}  \ ,
\end{align}
with $m_4=m-4c_1 M^2-2e_m M^4=m+m_{\rm tree}$.
The power of $\hbar$ indicates the loop order. The symbols $\Delta m_N^{(1)}$ and $\Delta m_N^{(2)}$ denote the one-loop and two-loop chiral corrections, respectively. They are related to the nucleon self-energy by
\begin{align}
\Delta m_N^{(1)}&=\Sigma_N^{(1)}(m_4,m_4)  \ , \\
\Delta m_N^{(2)}&=\Sigma_N^{(1)}(m_4,m_4)\,\Sigma_N^{(1)\prime}(m_4,m_4)+\Sigma_N^{(2)}(m_4,m_4) \ ,
\end{align}
where $\Sigma_N^{(1)\prime}$ is the derivative of the one-loop self-energy. It should be noted that $m_4$ is adopted such that the contributions from the mass insertion diagrams are automatically incorporated~\cite{Schindler:2007dr}. 

The leading two-loop nucleon self-energy can be calculated diagram by diagram. Up to $\mathcal{O}(p^5)$, there are $19$ Feynman diagrams in total: $2$ trees, $5$ one-loop and $12$ two-loop diagrams. The chiral results of loop diagrams are expressed in terms of loop integrals multiplied by invariant Lorentz scalars. For two-loop corrections, one encounters two-loop integrals. In our calculation, there are only five independent two-loop propagators $\mathcal{D}_i$($i=1,2,\cdots,5$), since we only have five irreducible scalar products constructed from the integration momenta $\ell_1$, $\ell_2$ and the external one $p$. Hence, the generic definition of two-loop integrals can be written as
\begin{align}
I_{\nu_1\nu_2\nu_3\nu_4
\nu_5}=
\int\int\frac{{\rm d}^d\ell_1}{(i\pi^{d/2})}
\frac{{\rm d}^d\ell_2}{(i\pi^{d/2})}
\frac{1}{\mathcal{D}_1^{\nu_1}\mathcal{D}_2^{\nu_2}\mathcal{D}_3^{\nu_3}\mathcal{D}_4^{\nu_4}\mathcal{D}_5^{\nu_5}} \ , 
\end{align}
where $\nu_{1,\cdots, 5}$ are integers. If all the $\nu$ indices are non-negative, the integral is named scalar integrals. If there exists at least a negative $\nu$, the integral is called tensor integrals. All the involved integrals can be reduced to a set of master integrals via the procedure of integration by parts~\cite{Weinzierl:2022eaz,Wu:2025aeg}. In our case, the set of master integrals contains $3$ one-loop and $10$ two-loop integrals. The master integrals can be either analytically solved by using the differential equations~\cite{Kotikov:1990kg,Henn:2013pwa} or numerically computed by e.g. \texttt{AMFlow}~\cite{Liu:2022chg} or \texttt{AmpRed}~\cite{Chen:2025paq}.
 
The renormalization of BChPT at two-loop order is non-trivial at all, in particular the notable PC breaking also emerges when the matter fields (here baryon) are massive. The application of EOMS beyond one-loop is discussed in e.g . Refs.~\cite{Schindler:2006ha,Schindler:2007dr, Conrad:2024phd, Conrad:2024sla, Liang:2025cjd, Chen:2024twu}, and a detailed illustration can be found in Ref.~\cite{Liang:2025cjd}.
The EOMS scheme is a two-step renormalization approach: the subtraction of the ultraviolet (UV) divergences in dimensional regularization followed by a finite shift canceling the PC breaking terms.  To remove the non-local UV divergences, the bare low-energy constants (LECs) in the one-loop diagram are split in the following way,
\begin{align}
X=-\frac{\beta_X^{(1)}}{\epsilon\Lambda}+X^r +\mathcal{O}(\epsilon)\ ,\quad X\in\{m, g, c_1,c_2,c_3, \ell_3, \ell_4, d_{16}, d_{18}\} 
\end{align}
with $\beta_X^{(1)}$ being one-loop beta functions and $\Lambda=1/(16\pi^2)$. To get rid of the local UV divergences, the bare LECs in the tree diagrams are split as
\begin{align}
X=\frac{\beta_X^{(22)}}{\epsilon^2\Lambda^2}
-\frac{\beta_X^{(21)}}{\epsilon\Lambda^2}-\frac{\beta_X^{(11)}}{\epsilon\Lambda}+X^r\ ,\quad X\in\{m, c_1, e_m\} \ ,
\end{align}
where $\beta_X^{(ij)}$ ($i,j\in\{1,2\}$) denote two-loop beta functions~\cite{Liang:2025cjd}.

It is well known that BChPT calculations suffer from the PC breaking problem~\cite{Gasser:1987rb}, due to the nucleon mass being nonzero in the chiral limit. In order to identify the PC breaking terms, we impose the method of dimensional counting analysis (DCA)~\cite{Gegelia:1999qt}, equivalent to the strategy of regions~\cite{Smirnov:2002pj}. With the help of DCA, three types of the PC breaking terms can be found. The first type corresponds to non-local PC breaking terms, which are canceled by the one-loop sub-diagrams via  
\begin{align}
X^r=\widetilde{X}+\delta_X^{(1)}\ ,\quad X\in\left\{g, m, c_1, c_2, c_3\right\} \ . 
\end{align}
The other two types correspond to one-loop and two-loop local PC breaking terms, respectively. Both of them are absorbed by the tree-level counterterms through the following splitting of LECs,
\begin{align}    
X^r=\widetilde{X}+\delta_X^{(21)}+\delta_X^{(22)} \ , \quad X\in\left\{m, c_1, e_m \right\} \ .
\end{align}
It should be mentioned that the PC-breaking-term subtraction is carried out in $d$-dimensional space-time. For more details of the calculation, see Ref.~\cite{Liang:2025cjd}.

Finally, we obtain the chiral result of nucleon mass up to $\mathcal{O}(p^5)$ within the full EOMS scheme,
\begin{align}
m_N=\widetilde{m}-\underbrace{4 \widetilde{c}_1 M^2}_{\mathcal{O}(p^2)} + \underbrace{\bar{m}_N^{(1a)}}_{\mathcal{O}(p^3)} -\underbrace{2 \widetilde{e}_m M^4 +\bar{m}_N^{(1b)}}_{\mathcal{O}(p^4)} + \underbrace{\bar{m}_N^{(1c)}+ 2 \bar{m}_N^{(1d)} + \bar{m}_N^\text{2-loop}+\bar{m}_N^\text{sub-diag.}}_{\mathcal{O}(p^5)}  \ , 
\end{align}
The quantities with over bars mean that the involved UV divergences and PC breaking terms are subtracted. A tilde stands for the parameter being EOMS-renormalized. Explicit expressions of each term are shown in Ref.~\cite{Liang:2025cjd}. This result possesses the following merits. It is renormalization scale independent, respects the original analytical property, and maintains correct power counting.

With the full EOMS result, one can further perform chiral expansion in powers of pion mass, which can be clearly formulated as
\begin{align}
m_N&=m+k_1 M^2+k_2 M^3+k_3 M^4\ln\frac{M}{\mu}+k_4 M^4+k_5 M^5\ln\frac{M}{\mu}+k_6 M^5 \ .
\end{align}
Explicit expressions of the coefficients, $k_1$ to $k_6$, can be found in Refs.~\cite{Liang:2025cjd, Chen:2024twu}. Nevertheless, one should notice that the chiral expansion unavoidably enforces a truncation at a certain order, which changes the analytic property of the original expression. Likewise, we can obtain the chiral expression in the IR prescription, which is identical to the one derived in Refs.~\cite{Schindler:2006ha,Schindler:2007dr}. Note that $k_3$ and $k_5$ are scheme independent, guaranteed by the renormalization group equations~\cite{Bijnens:2014ila}.

\section{Chiral extrapolation\label{sec3}}

\begin{figure}[ht]
\centering
\includegraphics[width=0.48\linewidth]{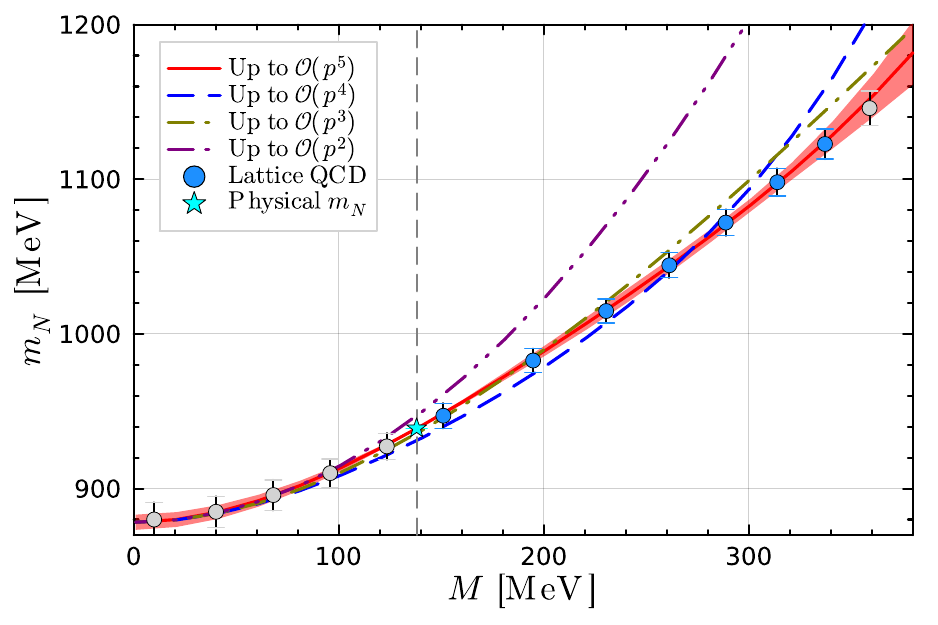}
\includegraphics[width=0.48\linewidth]{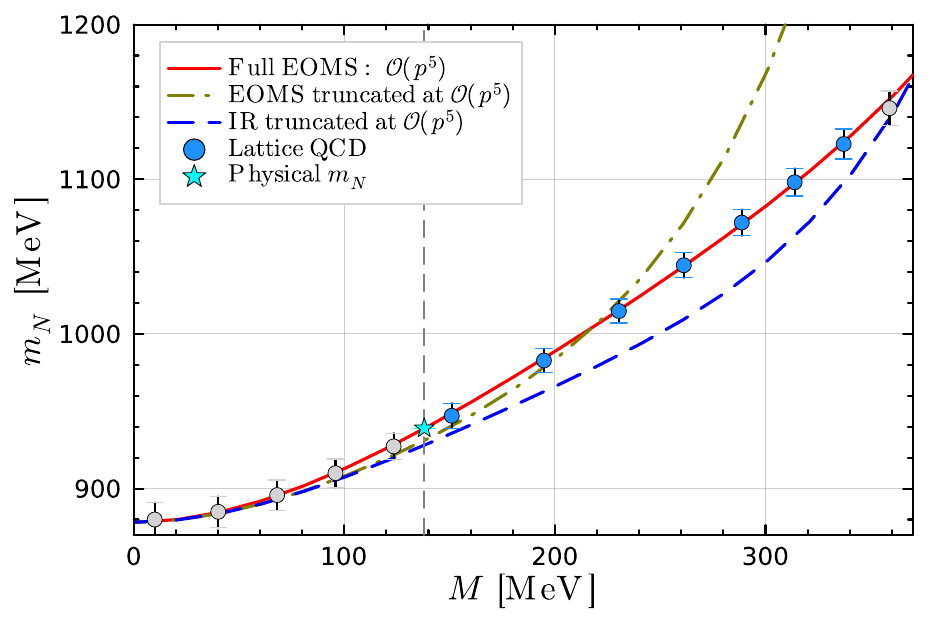}
\caption{
Left: The $M$-dependence of the nucleon mass at $\mathcal{O}(p^5)$. 
Right: Comparison of the nucleon mass calculated using: (a) full EOMS, (b) EOMS truncated at $\mathcal{O}(p^5)$, and (c) IR truncated at $\mathcal{O}(p^5)$.
The dashed gray line indicates the physical poin mass and the physical nucleon mass is marked by the cyan star~\cite{Liang:2025cjd}.
}
\label{fig:full_mass}
\end{figure}

The two-loop chiral result, discussed in the previous section, enables us to perform chiral extrapolation of recent lattice QCD data, which is displayed in the left panel of Fig.~\ref{fig:full_mass}. The $\mathcal{O}(p^5)$ nucleon mass curve exhibits excellent agreement with lattice QCD data~\cite{Hu:2024mas} across a wide range of pion masses, even beyond the fitting region. The uncertainty band reflects the variation of the fitted parameters within their $1\sigma$ uncertainties. For comparison, pion mass dependence of the nucleon mass at $\mathcal{O}(p^2)$, $\mathcal{O}(p^3)$ and $\mathcal{O}(p^4)$ is also shown. At the physical pion mass, the nucleon mass reads
\begin{align}
m_N=\left\{878.2+68.8+[-11.4]+[-4.6]+7.9 \right\}
~\rm{MeV}
=938.9~\rm{MeV} \ .
\label{eq:mass}
\end{align}
The net one-loop contribution is $(-11.4-4.6)=-16$~MeV, larger than the $\mathcal{O}(p^5)$ two-loop contribution $\sim 7.9$~MeV. In this sense, the chiral series of the nucleon mass converges well. A comparison of our full-EOMS, EOMS-truncated, and IR-truncated results is given in the right panel of the Fig.~\ref{fig:full_mass} using the same set of the LECs. As expected, they agree well with each other for small pion masses but start to deviate as the pion mass increases.

\section{Summary and outlook\label{sec4}}
The calculation of the nucleon mass in BChPT using the EOMS scheme is reviewed. The feasibility of the EOMS scheme at the two-loop level is demonstrated in detail: both the UV divergences and the notable PC breaking terms can be properly absorbed by LECs in the one-loop and tree-level counter terms. It is found that the $\mathcal{O}(p^5)$ contribution in the full EOMS scheme is small, implying a good convergency property at the two-loop level. We also show that the EOMS-truncated and IR-truncated results are not suitable for chiral extrapolation due to the fact that the analytic structure is changed by the operation of chiral truncation. This successful two-loop representation of the nucleon mass marks a significant milestone in BChPT, opening a promising era of two-loop BChPT to explore the nucleon property with high accuracy.

\section{Acknowledgments}
This work is supported by the Science Foundation of Hebei Normal University with Contract No.~L2025B09 and No.~L2023B09; by Science Research Project of Hebei Education Department under Contract No.~QN2025063;  by National Nature Science Foundations of China (NSFC) under Contract No.~12275076, No.~11905258, No.~12335002, No.~12447101, No.~12475078, and No.~12125507; by the Science Fund for Distinguished Young Scholars of Hunan Province under Grant No.~2024JJ2007; by the Chinese Academy of Sciences under Grant No.~YSBR-101.


\begin{thebibliography}{99}
%%%%%%%%%%%%%%%%%%
%\cite{Ji:1994av}
\bibitem{Ji:1994av}
X.~D.~Ji,
%``A QCD analysis of the mass structure of the nucleon,''
Phys. Rev. Lett. \textbf{74}, 1071-1074 (1995).
%doi:10.1103/PhysRevLett.74.1071
%[arXiv:hep-ph/9410274 [hep-ph]].
%275 citations counted in INSPIRE as of 31 Jul 2025

%\cite{Lorce:2017xzd}
\bibitem{Lorce:2017xzd}
C.~Lorc{\'e},
%``On the hadron mass decomposition,''
Eur. Phys. J. C \textbf{78}, no.2, 120 (2018).
%doi:10.1140/epjc/s10052-018-5561-2
%[arXiv:1706.05853 [hep-ph]].
%117 citations counted in INSPIRE as of 30 Jul 2025

%\cite{Bottino:1999ei}
\bibitem{Bottino:1999ei}
A.~Bottino, F.~Donato, N.~Fornengo and S.~Scopel,
%``Implications for relic neutralinos of the theoretical uncertainties in the neutralino nucleon cross-section,''
Astropart. Phys. \textbf{13}, 215-225 (2000).
%doi:10.1016/S0927-6505(99)00122-X
%[arXiv:hep-ph/9909228 [hep-ph]].
%184 citations counted in INSPIRE as of 11 Jul 2025

%\cite{Feynman:1939zza}
\bibitem{Feynman:1939zza}
R.~P.~Feynman,
%``Forces in Molecules,''
Phys. Rev. \textbf{56}, 340-343 (1939).
%doi:10.1103/PhysRev.56.340
%455 citations counted in INSPIRE as of 18 Jul 2025

\bibitem{Hellmann:1937book}%Hans 
H. Hellmann, Einführung in die Quantenchemie,
Springer Spektrum Berlin, Heidelberg(1937).

%\cite{McGovern:1998tm}
\bibitem{McGovern:1998tm}
J.~A.~McGovern and M.~C.~Birse,
%``On the absence of fifth order contributions to the nucleon mass in heavy baryon chiral perturbation theory,''
Phys. Lett. B \textbf{446}, 300-305 (1999).
%doi:10.1016/S0370-2693(98)01550-0
%[arXiv:hep-ph/9807384 [hep-ph]].
%58 citations counted in INSPIRE as of 11 Jul 2025

%\cite{Schindler:2006ha}
\bibitem{Schindler:2006ha}
M.~R.~Schindler, D.~Djukanovic, J.~Gegelia and S.~Scherer,
%``Chiral expansion of the nucleon mass to order(q**6),''
Phys. Lett. B \textbf{649}, 390-393 (2007).
%doi:10.1016/j.physletb.2007.04.034
%[arXiv:hep-ph/0612164 [hep-ph]].
%45 citations counted in INSPIRE as of 11 Jul 2025

%\cite{Schindler:2007dr}
\bibitem{Schindler:2007dr}
M.~R.~Schindler, D.~Djukanovic, J.~Gegelia and S.~Scherer,
%``Infrared renormalization of two-loop integrals and the chiral expansion of the nucleon mass,''
Nucl. Phys. A \textbf{803}, 68-114 (2008)
[erratum: Nucl. Phys. A \textbf{1010}, 122175 (2021)].
%doi:10.1016/j.nuclphysa.2008.01.023
%[arXiv:0707.4296 [hep-ph]].
%37 citations counted in INSPIRE as of 11 Jul 2025

%\cite{Becher:1999he}
\bibitem{Becher:1999he}
T.~Becher and H.~Leutwyler,
%``Baryon chiral perturbation theory in manifestly Lorentz invariant form,''
Eur. Phys. J. \textbf{9}, no.4, 643-671 (1999).
%doi:10.1007/PL00021673
%[arXiv:hep-ph/9901384 [hep-ph]].
%636 citations counted in INSPIRE as of 22 Jul 2025

%\cite{Conrad:2024phd}
\bibitem{Conrad:2024phd}
N.~D.~Conrad, A.~Gasparyan and E.~Epelbaum,
%``Two-loop calculation of the nucleon self-energy,''
PoS \textbf{CD2021}, 074 (2024).
%doi:10.22323/1.413.0074
%3 citations counted in INSPIRE as of 11 Jul 2025

%\cite{Conrad:2024sla}
\bibitem{Conrad:2024sla}
N.~D.~Conrad, A.~M.~Gasparyan and E.~Epelbaum,
%``Nucleon Self-Energy Including Two-Loop Contributions,''
EPJ Web Conf. \textbf{303}, 02001 (2024).
%doi:10.1051/epjconf/202430302001
%2 citations counted in INSPIRE as of 11 Jul 2025

%\cite{Chen:2024twu}
\bibitem{Chen:2024twu}
L.~B.~Chen, S.~Hu, Y.~Jia and Z.~Mo,
%``Light quark mass dependence of nucleon mass to two-loop order,''
[arXiv:2406.04124 [hep-ph]].
%2 citations counted in INSPIRE as of 11 Jul 2025

%\cite{Liang:2025cjd}
\bibitem{Liang:2025cjd}
Z.~R.~Liang, H.~X.~Chen, F.~K.~Guo, Z.~H.~Guo and D.~L.~Yao,
%``Chiral representation of the nucleon mass at leading two-loop order,''
JHEP \textbf{04}, 192 (2025).
%doi:10.1007/JHEP04(2025)192
%[arXiv:2502.19168 [hep-ph]].
%2 citations counted in INSPIRE as of 11 Jul 2025

%\cite{Ren:2016aeo}
\bibitem{Ren:2016aeo}
X.~L.~Ren, L.~Alvarez-Ruso, L.~S.~Geng, T.~Ledwig, J.~Meng and M.~J.~Vicente Vacas,
%``Consistency between SU(3) and SU(2) covariant baryon chiral perturbation theory for the nucleon mass,''
Phys. Lett. B \textbf{766} (2017), 325-333.
%doi:10.1016/j.physletb.2017.01.024
%[arXiv:1606.03820 [nucl-th]].
%11 citations counted in INSPIRE as of 31 Jul 2025

%\cite{Yao:2016vbz}
\bibitem{Yao:2016vbz}
D.~L.~Yao, D.~Siemens, V.~Bernard, E.~Epelbaum, A.~M.~Gasparyan, J.~Gegelia, H.~Krebs and U.~G.~Mei{\ss}ner,
%``Pion-nucleon scattering in covariant baryon chiral perturbation theory with explicit Delta resonances,''
JHEP \textbf{05}, 038 (2016).
%doi:10.1007/JHEP05(2016)038
%[arXiv:1603.03638 [hep-ph]].
%104 citations counted in INSPIRE as of 31 Jul 2025

%\cite{Liang:2023scp}
\bibitem{Liang:2023scp}
Z.~R.~Liang, P.~C.~Qiu and D.~L.~Yao,
%``One-loop analysis of the interactions between doubly charmed baryons and Nambu-Goldstone bosons,''
JHEP \textbf{07}, 124 (2023).
%doi:10.1007/JHEP07(2023)124
%[arXiv:2303.03370 [hep-ph]].
%4 citations counted in INSPIRE as of 31 Jul 2025


%\cite{Weinzierl:2022eaz}
\bibitem{Weinzierl:2022eaz}
S.~Weinzierl,
%``Feynman Integrals. A Comprehensive Treatment for Students and Researchers,''
Springer, 2022,
ISBN 978-3-030-99557-7.%, 978-3-030-99560-7, 978-3-030-99558-4.
%doi:10.1007/978-3-030-99558-4
%[arXiv:2201.03593 [hep-th]].
%196 citations counted in INSPIRE as of 25 Jul 2025

%\cite{Wu:2025aeg}
\bibitem{Wu:2025aeg}
Z.~Wu, J.~B{\"o}hm, R.~Ma, J.~Usovitsch, Y.~Xu and Y.~Zhang,
%``Performing integration-by-parts reductions using NeatIBP 1.1 + Kira,''
[arXiv:2502.20778 [hep-ph]].
%8 citations counted in INSPIRE as of 31 Jul 2025

%\cite{Kotikov:1990kg}
\bibitem{Kotikov:1990kg}
A.~V.~Kotikov,
%``Differential equations method: New technique for massive Feynman diagrams calculation,''
Phys. Lett. B \textbf{254}, 158-164 (1991).
%doi:10.1016/0370-2693(91)90413-K
%1051 citations counted in INSPIRE as of 29 Jul 2025

%\cite{Henn:2013pwa}
\bibitem{Henn:2013pwa}
J.~M.~Henn,
%``Multiloop integrals in dimensional regularization made simple,''
Phys. Rev. Lett. \textbf{110}, 251601 (2013).
%doi:10.1103/PhysRevLett.110.251601
%[arXiv:1304.1806 [hep-th]].
%1048 citations counted in INSPIRE as of 29 Jul 2025

%\cite{Liu:2022chg}
\bibitem{Liu:2022chg}
X.~Liu and Y.~Q.~Ma,
%``AMFlow: A Mathematica package for Feynman integrals computation via auxiliary mass flow,''
Comput. Phys. Commun. \textbf{283}, 108565 (2023).
%doi:10.1016/j.cpc.2022.108565
%[arXiv:2201.11669 [hep-ph]].
%258 citations counted in INSPIRE as of 29 Jul 2025

%\cite{Chen:2025paq}
\bibitem{Chen:2025paq}
W.~Chen,
%``An Improvement of AmpRed: Analytic Continuation of Complex Integrals,''
[arXiv:2505.13540 [hep-ph]].
%0 citations counted in INSPIRE as of 31 Jul 2025

%\cite{Gasser:1987rb}
\bibitem{Gasser:1987rb}
J.~Gasser, M.~E.~Sainio and A.~Svarc,
%``Nucleons with chiral loops,''
Nucl. Phys. B \textbf{307}, 779-853 (1988).
%doi:10.1016/0550-3213(88)90108-3
%1034 citations counted in INSPIRE as of 22 Jul 2025

%\cite{Gegelia:1999qt}
\bibitem{Gegelia:1999qt}
J.~Gegelia, G.~Japaridze and X.~Q.~Wang,
%``Is Heavy baryon approach necessary?,''
J. Phys. G \textbf{29}, 2303-2309 (2003).
%doi:10.1088/0954-3899/29/9/322
%[arXiv:hep-ph/9910260 [hep-ph]].
%74 citations counted in INSPIRE as of 11 Jul 2025

%\cite{Smirnov:2002pj}
\bibitem{Smirnov:2002pj}
V.~A.~Smirnov,
%``Applied asymptotic expansions in momenta and masses,''
Springer Tracts Mod. Phys. \textbf{177}, 1-262 (2002).
%328 citations counted in INSPIRE as of 25 Jul 2025

%\cite{Bijnens:2014ila}
\bibitem{Bijnens:2014ila}
J.~Bijnens and A.~A.~Vladimirov,
%``Leading logarithms for the nucleon mass,''
Nucl. Phys. B \textbf{891}, 700-719 (2015).
%doi:10.1016/j.nuclphysb.2014.12.015
%[arXiv:1409.6127 [hep-ph]].
%9 citations counted in INSPIRE as of 11 Jul 2025

%\cite{Hu:2024mas}
\bibitem{Hu:2024mas}
B.~Hu, H.~Du, X.~Jiang, K.~F.~Liu, P.~Sun and Y.~B.~Yang,
%``Unveiling the Strong Interaction origin of Baryon Masses with Lattice QCD,''
[arXiv:2411.18402 [hep-lat]].
%6 citations counted in INSPIRE as of 21 Jul 2025

\end{thebibliography}
\end{document}